\documentclass[twocolumn,preprintnumbers]{revtex4}
\usepackage{epsf}

\begin{document}
\preprint{KUNS-2334}
\title{Proton dominance in the $2^+_2 \rightarrow 0^+_1$ transition of $N = Z\pm 2$ nuclei around $^{28}$Si}
\author{Yoshiko Kanada-En'yo}
\affiliation{Department of Physics, Kyoto University, Kyoto 606-8502, Japan}
\begin{abstract}
$E2$ transitions in  $^{30}$Si are investigated in relation with intrinsic deformations
based on a method of antisymmetrized molecular dynamics.
By comparing $E2$ transition strengths in the mirror nuclei $^{30}$Si and $^{30}$S,
transition matrix amplitudes $M_p$ and $M_n$ 
for protons and neutrons are discussed in mirror analysis.
Particular attention is paid to the 
$M_n/M_p$ ratio in the transition from the $2^+_2$ state to the $0^+_1$ state. 
The $M_n/M_p$ ratio in $^{26}$Mg and $^{26}$Si is also investigated.  
It is found that the proton dominance in the transition $2^+_2\rightarrow 0^+_1$
in $^{30}$Si and $^{26}$Si originates in the oblate trend of the $Z=14$ proton structure.
\end{abstract}
\maketitle

\noindent

\section{Introduction} \label{ref:introduction}
Nuclear systems often show intrinsic 
quadrupole deformations. 
In many nuclei except for shell-closed nuclei,  
normal deformations are found in the ground bands. 
Shell effects are essential for the deformations
as described by the Nilsson model, and therefore, 
a variety of shapes appear depending on the mass number
particular in mid-shell nuclei.
In $Z=N$ nuclei in the $sd$-shell region, a nuclear shape 
rapidly changes as a function of $Z=N$ reflecting the proton-neutron coherent shell effects 
in deformed systems.
For instance, the prolate ground band 
in $^{20}$Ne and the oblate one in $^{28}$Si are known.
These shapes are easily understood from the shell effects in the Nilsson model, 
where the $Z=N=10$ and $Z=N=14$ shell gaps appear in the prolate and the oblate deformations,  
respectively.

For $Z\ne N$ nuclei, deformation phenomena are not so simple as in $Z=N$ nuclei.  
If the shell effect for proton orbits and that for neutron ones compete with each other,
the proton shape can be affected by the neutron structure, or decoupling between proton and neutron 
deformations may occur. 
The latter case might be possible in light-mass nuclei and it may be observed in 
the quadrupole transition properties such as the ratio of 
the neutron transition matrix amplitude to the proton one (so-called $M_n/M_p$ ratio).
Such the decoupling between proton and neutron shapes has been suggested, for instance, in 
$^{16}$C for which the enhanced $M_n/M_p$ ratio, i.e., the neutron dominance was observed 
in the ground-band transition \cite{Elekes04}. The neutron dominance 
was described by the oblate proton and prolate neutron shapes 
\cite{KanadaEn'yo:2004bi,Takashina:2005bs}.
One of the reasons for the characteristic structure of $^{16}$C 
is that a $Z=6$ system favors an oblate proton deformation
while a $N=10$ nucleus has the prolate trend of the neutron deformation 
because of each shell effect of the proton and neutron parts.
To clarify the oblate tendency of the proton structure in $^{16}$C,
a possible $K=2$ side band and its transition properties should be experimentally observed.
In general, 
such a nucleus having oblate proton and prolate neutron structures may show an
isovector triaxiality. If the case, in addition to the $K=0$ band,
a $K=2$ side band is constructed with the rotation
around the symmetric axis of the prolate neutron part.
Since the proton contribution should be dominant while the neutron contribution 
is minor for this rotation, the transition from the $2^+_2$ state in the $K=2$ band 
to the ground state may show the proton dominance resulting in a small $M_n/M_p$ ratio.
Although the $K=2$ side band was theoretically suggested,  
the $2^+_2$ state in $^{16}$C has not been experimentally confirmed yet, unfortunately. 
$^{10}$C is another candidate for the nuclei with the isovector 
triaxiality\cite{KanadaEn'yo:2004bi,KanadaEn'yo:1996hi}, but the 
transition strength from the $2^+_2$ state to the $0^+_1$ state has not yet been measured.  

A similar situation of the isovector triaxiality is expected in $Z=14$ nuclei, i.e., Si isotopes 
because the oblate proton structure is favored due to the $Z=14$ shell effect. 
Let us consider deformations of $^{26}$Si and $^{30}$Si. 
From the negative sign of the $Q$-moments for the $2^+_1$ states 
in the $Z=N=12$ and $Z=N=16$ systems, 
the prolate trend of neutron structure is expected in $N=12$ and $N=16$ systems.
Considering the prolate tendency of the neutron part, 
the isovector triaxiality in $^{26}$Si and $^{30}$Si would be possible  
and it might lead to the proton dominance in the $2^+_2\rightarrow 0^+_1$ transition
resulting in a small $M_n/M_p$ ratio.
For $^{26}$Si and $^{30}$Si,
the $M_n/M_p$ ratios for the $2^+_2\rightarrow 0^+_1$ transition
as well as the $2^+_1\rightarrow 0^+_1$ was
experimentally determined by mirror analysis\cite{Alexander82,Alexander82b} and inelastic 
scattering data\cite{Sciuccati85}. 
The reported values of the $M_n/M_p$ ratio in the $2^+_2\rightarrow 0^+_1$ 
are $0.50\pm 0.07$ and $0.52\pm 0.03$ for $^{26}$Si and $^{30}$Si, respectively, and
they indicate the proton dominance.
The data of neutron and proton transition matrices, $M_n$ and $M_p$, in this mass region 
are qualitatively reproduced by shell model calculations\cite{Brown:1982zz}
and they were discussed in relation with effective charges coming from core polarization
\cite{Brown:1982zz,Alexander85}. The proton dominance in the 
$2^+_2\rightarrow 0^+_1$ of $^{26}$Si was discussed also in relation with the $\gamma$ 
softness in the calucaltions based on the quadrupole collective 
Hamiltonian\cite{Hinohara:2010sf}.

Our aim is to understand the properties of 
proton and neutron transition matrices from the viewpoint of 
deformations of proton and neutron parts. In particular, the proton dominance 
in the $2^+_2\rightarrow 0^+_1$ is a focusing feature which might be interpreted 
in connection to decoupling of proton and neutron shapes. 
For this aim, we apply a theoretical approach of antisymmetrized molecular dynamics(AMD).
The AMD has been successfully applied for study of structures of $p$-shell and $sd$-shell 
nuclei\cite{ENYObc,ENYOsup,AMDrev}.
The present method of AMD calculations is the same as those used for investigation of 
deformation phenomena in $^{28}$Si, $^{24}$Ne, $^{22}$O, and $^{20}$C\cite{KanadaEn'yo:2004cv} and
those in C isotopes\cite{KanadaEn'yo:2004bi}. 

Applying the AMD method, we investigate the transition properties of $^{30}$Si.
Assuming the mirror symmetry, the calculated proton and neutron transition matrix amplitudes 
are compared with the data evaluated by the observed $E2$ transition strength for
$^{30}$Si and $^{30}$S, and they are discussed in connection with proton and neutron deformations.
Transition properties in $^{26}$Mg and $^{26}$Si are also discussed.
The proton dominance in the transition, $2^+_2\rightarrow 0^+_1$, for 
$^{10}$C and $^{16}$C is also shown as well as that for Si isotopes.

The paper is organized as follows. 
In the next section, the formulation of the present calculation
is explained. In \ref{sec:results}, The results for $^{30}$Si and $^{30}$S and those for $^{26}$Mg and $^{26}$Si 
are shown and the proton dominance of $2^+_2\rightarrow 0^+_1$ in $^{30}$Si and $^{26}$Si 
is discussed. The proton dominance in $^{10}$C and $^{16}$C is shown in \ref{sec:results2}.
In \ref{sec:summary}, a summary is given.

\section{Formulation}\label{sec:formulation}

Here we briefly explain the formulation of the present calculations.
Details of the formulation of AMD methods for nuclear structure 
study are explained in Refs.~\cite{ENYObc,ENYOsup,AMDrev}.
The method of the present calculations is basically same as that
in Refs.~\cite{ENYObc,KanadaEn'yo:2004cv}.

An AMD wave function $\Phi_{\rm AMD}$ for a system with the mass number $A$ 
is given by a single Slater determinant of Gaussian wave packets as,
\begin{equation}
 \Phi_{\rm AMD}({\bf Z}) = \frac{1}{\sqrt{A!}} {\cal{A}} \{
  \varphi_1,\varphi_2,...,\varphi_A \},
\end{equation}
where the $i$-th single-particle wave function is written as,
\begin{eqnarray}
 \varphi_i&=& \phi_{{\bf X}_i}\chi_i\tau_i,\\
 \phi_{{\bf X}_i}({\bf r}_j) &\propto& 
\exp\bigl\{-\nu({\bf r}_j-\frac{{\bf X}_i}{\sqrt{\nu}})^2\bigr\},
\label{eq:spatial}\\
 \chi_i &=& (\frac{1}{2}+\xi_i)\chi_{\uparrow}
 + (\frac{1}{2}-\xi_i)\chi_{\downarrow}.
\end{eqnarray}
Here the isospin function $\tau_i$ is fixed to be up(proton) or down(neutron).
The orientation of intrinsic spin 
$\xi_i$ is also fixed to be $1/2$ or $-1/2$ in the present calculations as done in Refs.\cite{ENYObc,KanadaEn'yo:2004cv}.
The width parameter for Gaussian wave packet is taken to be $\nu=0.15$ fm$^{-2}$ which 
is the optimum value for $^{28}$Si used in Ref.~\cite{KanadaEn'yo:2004cv}.
The spatial part, $\phi_{{\bf X}_i}$, is written by 
a Gaussian wave packet localized at the certain position 
${\bf X}_i$ in the phase space. Then, the AMD wave function 
is expressed by $\{{\bf X}_i\}$ which indicate the Gaussian centers for 
all the single-particle wave functions and are treated as the 
independent complex variational parameters.

We perform energy variation for a parity-eigen state, 
$P^\pm\Phi_{\rm AMD}\equiv \Phi^\pm_{\rm AMD}$, 
projected from an AMD wave function
by means of the frictional cooling method\cite{ENYObc}.
We consider the AMD wave function obtained by the energy variation 
as an intrinsic state, and 
total-angular-momentum projection($P^J_{MK}$) is 
performed after the variation to calculate such observables
as energies and transition strengths.
The $K$-mixing is incorporated in the total-angular-momentum projection 
for non-zero $J$ states.

As shown later, two local minimum solutions are obtained in the energy variation 
for the $Z=14$ systems. The two minima almost degenerate
to each other and may correspond to shape coexistence phenomena
as already discussed in the $N=14$ systems\cite{KanadaEn'yo:2004cv}.
Comparing the calculated structure properties such as transition strengths 
with the experimental data, we assign one of two minima to the intrinsic state of the 
ground band. 

By using the obtained wave functions, we calculate the transition matrix amplitudes $M_p$ and $M_n$ for
proton and neutron, respectively, and also the $E2$ transition strengths. They are defined as follows.
\begin{eqnarray}
M_p&=& \langle f || P(t_z=1/2) r Y^2_\mu|| i \rangle, \\ 
M_n&=& \langle f || P(t_z=-1/2) r Y^2_\mu|| i \rangle, \\ 
B(E2)&=& \frac{e^2}{2J_i+1} M_p^2. \\  
\end{eqnarray}
Here $P(t_z=\pm 1/2)$ are the isospin projection operators for protons and neutrons. 

\section{Results}\label{sec:results}
\subsection{Effective interaction}
The effective nuclear interactions adopted in the present work
consist of the central force, the
 spin-orbit force and Coulomb force.
We adopt MV1 force \cite{TOHSAKI} as the central force.
The MV1 force contains a zero-range three-body force 
in addition to the two-body interaction.
For interaction parameters, we use the same parameter set as that in Ref.~\cite{KanadaEn'yo:2004cv}.
Namely, the MV1 force (case 1) with the parameters 
$b=0$, $h=0$, and $m=0.62$ is used. 
For the spin-orbit force, 
the two-range Gaussian form of the G3RS force \cite{LS} is adopted. 
The strengths of the spin-orbit force are $u_{I}=-u_{II}=2800$ MeV. 
These strengths were adjusted to reproduce energy levels of $^{28}$Si in Ref.~\cite{KanadaEn'yo:2004cv}.

\subsection{$^{30}$Si and $^{30}$S}

After the energy variation for parity projected AMD wave functions, 
two local minimum solutions (A) and (B) are obtained for $^{30}$Si.
The state (A) shows a smaller deformation as 
$(\beta_p,\gamma_p)=(0.18, 0.01\pi)$ and $(\beta_n,\gamma_n)=(0.20, 0.00\pi)$ 
and the state (B) has a lager deformation as
$(\beta_p,\gamma_p)=(0.35, 0.00\pi)$ and $(\beta_n,\gamma_n)=(0.26, 0.00\pi)$.
Here the definition of the quadrupole deformation parameters 
$(\beta_p,\gamma_p)$ for proton density distribution 
and $(\beta_n,\gamma_n)$ for neutron one is given in Ref.~\cite{KanadaEn'yo:1996hi}.

\begin{figure}[th]
\epsfxsize=0.46\textwidth
\centerline{\epsffile{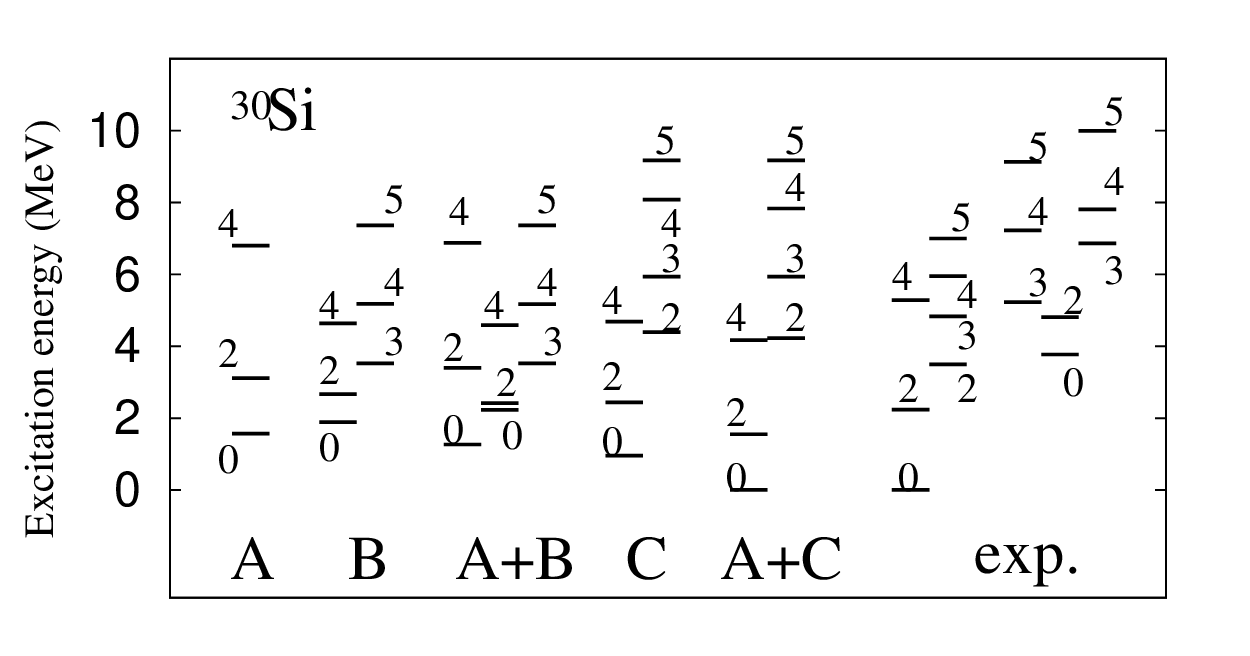}}
\caption{Energy levels of $^{30}$Si for positive parity states. The calculated levels are obtained 
by the projection from the state (A), the state (B), the superposition (A+B), the state (C), and
the superposition (A+C). The energies are measured from the $0^+_1$ calculated with (A+C).
The experimental levels are the positive parity bands
observed by the gamma-ray measurements in Ref.~\cite{Glatz80}.}
\label{fig:spe-si30fig}
\end{figure}

After the angular momentum projection, the energy levels are obtained from 
the states (A) and (B) (Fig.~\ref{fig:spe-si30fig}). 
The bands constructed from two intrinsic states, (A) and (B), almost degenerate. 
As shown later, the band (A) shows good agreements with experimental data such
as the $Q$ moment and $E2$ transition strengths  for
the ground band, and therefore, we can assign the band (A) to the
experimental ground band. The band (B) might correspond to the excited band, 
but the excitation energy seems to be underestimated in the present calculation.  
Because of this underestimation, 
the energy levels calculated by superposition of (A) and (B) show
the band mixing feature around $J\sim 2$ which is inconsistent with the experimental data. 
Since our interest is in the transition properties of the ground band and its side band,
we consider the state (A) for the ground band and omit the mixing with the state (B)  
in the following discussions. 

\begin{figure}[th]
\epsfxsize=0.4\textwidth
\centerline{\epsffile{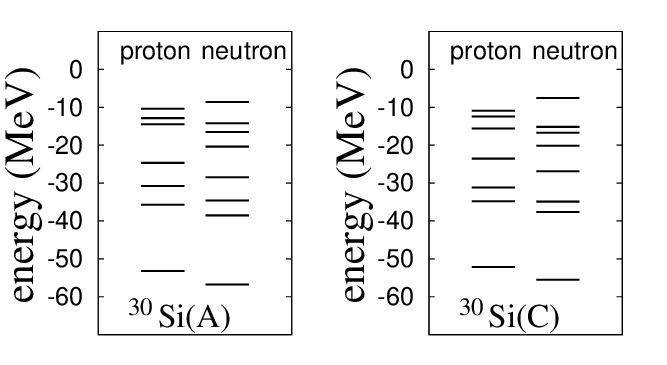}}
\caption{Single-particle energies in the intrinsic wave functions 
$^{30}$Si(A) and $^{30}$Si(C).}
\label{fig:hfe-si30}
\end{figure}

Experimentally, the $E2$ transition strengths were measured up to high spin states, and 
the $K^\pi=2^+$ side band by the $K^\pi=0^+$ ground band was
identified\cite{Glatz80}. 
In the present calculation, 
no $K^\pi=2^+$ side band member is obtained from the band (A) because of the
prolate intrinsic shape of the state (A).
Owing to the $Z=14$ shell gap in the oblate deformation, it is naturally expected that 
$^{30}$Si may be soft agaist $\gamma$ deformation toward the oblate region. 
Therefore, we construct another intrinsic state (C) by the alternative energy variation where
we vary the single-particle neutron wave functions
but  freeze the proton configuration so that it has the same proton structure as that
of the oblate ground state of $^{28}$Si. 
It corresponds to the energy variation with the constraint of 
the oblate proton structure.  Thus obtained state (C) has an triaxial neutron structure 
of $(\beta_n,\gamma_n)=(0.21, 0.15\pi)$ 
with the oblate proton deformation of $(\beta_p,\gamma_p)=(0.27, 0.28\pi)$.
The single-particle energy levels in the intrinsic wave functions for the states (A) and
(C) are shown in Fig.~\ref{fig:hfe-si30}. The derivation of the single-particle energies 
of AMD wave functions is described, for example, in Refs.~\cite{Dote:1997zz,AMDrev}.
In the state (C), the single-particle energy spectra are consistent with those for oblate systems
where the $N=14$ shell gap is clearly seen, while the $N=14$ shell gap vanishes in the state (A). 

After the angular momentum projection, the $K^\pi=0^+$ and the $K^\pi=2^+$ side bands are
generated from the intrinsic state (C). We should stress that the absolute energy of the 
$0^+_1$ projected from the state (C) is relatively lower than that from the state (A)
even though the intrinsic energy of the state (C) is higher than (A) before the projection. 
It is because the triaxial state gains more energy than the prolate state in the angular momentum projection.
Finally we superpose the state (A) and (C), and obtain the energy levels for the ground and side bands. 
The energy levels calculated by the superposition (A+C) correspond well to the experimental levels 
for the $K^\pi=0^+$ and the $K^\pi=2^+$ bands except for a slightly higher excitation energy 
for the $K^\pi=2^+$ band.
Considering that the $K^\pi=2^+$ band members are constructed mainly from the $|K|=2$ components of the 
intrinsic state (C) and the ground band can be described by the $K=0$ states of (C), 
these bands are considered to be the $K^\pi=0^+$ and the $K^\pi=2^+$ side bands projected from the 
triaxial intrinsic state (C). 
We here note that the wave functions of the $K^\pi=0^+$ components of the state (C) and (A)
have large overlap with each other.
Therefore, an alternative interpretation is possible. Namely, the ground band 
is regarded as the rotational band of the prolate state (A) and the $K=2$ side band is the $\gamma$ vibration band
on the top of the prolate state, which is taken into account by
the superposition of (A) and (C).

We calculate the quadrupole transition properties by the superposition (A+C) and discuss the 
neutron and proton contributions based on the mirror analysis. 
We first show the calculated values of $B(E2)$, the proton transition matrix amplitudes 
$M_p$, and electric quadrupole moments $Q$ in $^{30}$Si in comparison with 
the experimental data in Table \ref{tab:be2-si30}. 
The theoretical values are in good agreement with the experimental data except for the inter-band
transition $4^+_2\rightarrow 2^+_1$. 

\begin{figure}[th]
\epsfxsize=0.4\textwidth
\centerline{\epsffile{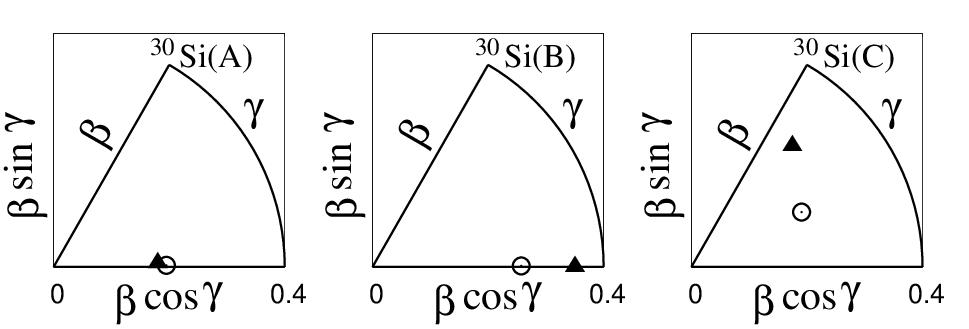}}
\caption{Deformation parameters $\beta$ and $\gamma$ of the intrinsic wave functions 
$^{30}$Si(A), $^{30}$Si(B), and $^{30}$Si(C). The filled triangles indicate $\beta_p$ and $\gamma_p$ for the proton part
and the open circles are $\beta_n$ and $\gamma_n$ for the neutron part}
\label{fig:defo-si30}
\end{figure}

We next discuss the $M_n/M_p$ ratio for the $2^+_1\rightarrow 0^+_1$ and $2^+_2\rightarrow 0^+_1$ transitions. 
We assume here the mirror symmetry that the neutron transition matrix amplitudes $M_n$ in $^{30}$Si is 
consistent with the proton transition matrix amplitudes $M_p$ in $^{30}$S.
Then, the experimental $M_n$ values for $^{30}$Si are evaluated by $B(E2)$ in $^{30}$S.
As shown in Table \ref{tab:mirror}, the experimental $M_n/M_p$ ratio for $2^+_1\rightarrow 0^+_1$  in 
$^{30}$Si is close to a unit
indicating that the proton and neutron parts equarly contribute to the ground-band transition.
What is striking is that the $M_n/M_p$ ratio for $2^+_2\rightarrow 0^+_1$ is 0.5 in the experimental data. 
This indicates the dominant proton matrix amplitude, which is about twice of the neutron one
in the transition from the side band to the ground state. 
The present calculation reproduces this trend of the quenched $M_n/M_p$ ratio, though they much 
underestimate the $M_n$ value for $2^+_2\rightarrow 0^+_1$ in $^{30}$Si.
The origin of the dominant proton contribution and the minor neutron one in this transition, 
$2^+_2\rightarrow 0^+_1$, 
is understood by the difference between proton and neutron structures. 
As mentioned before, the side band members are constructed mainly from the $|K|=2$ components of the 
intrinsic state (C), while the ground band is also approximately written by the $|K|=0$ states projected from the
state (C). 
That is, the $2^+_2$ state can be interpreted as the triaxial side band. 
As shown in Fig.~\ref{fig:defo-si30}, the state (C) shows the larger triaxiality of the proton shape than the neutron part,
and therefore, proton contribution should be dominant but neutron one is minor in the 
$2^+_2\rightarrow 0^+_1$. 

We should comment again that the $K=0$ components of the triaxial state (C) have large overlaps with those of 
the prolate state (A). In fact, the overlap between the $0^+$ states projected from (A) and (C) is 55\%.  
This is because the deformation parameter $\gamma$ is
not a coordinate but just an expectation value, and two states with different $\gamma$ values are not orthogonal
to each other. 
Even if a rotational band is constructed by the angular-momentum projection from an intrinsic state, 
the intrinsic shape is not observable but an interpretation which is useful to interprete the microscopic wave functions
projected from the intrinsic state as the rotational band members. 
Therefore, strictly speaking, it is not easy to clearly distinguish between rotational $K=2$ modes 
of a static triaxial state and $\gamma$-vibrational modes in such the system. 
Then, we can consider the alternative interpretation for the $2^+_2$ state as 
the $\gamma$ vibration on the prolate $K=0$ ground band. 
Again, the proton dominance can be easily understood by the 
$\gamma$ vibration of the proton structure. 
In any cases, we can conclude that the proton dominance of the transition, $2^+_2\rightarrow 0^+_1$, 
in $^{30}$Si originates in the $\gamma$ softness of the proton part in the $Z=14$ system,
and the difference between proton and neutron structures is essential.

\begin{table}[ht]
\caption{\label{tab:be2-si30}
$E2$ transition strengths of the in-band and inter-band transitions
in $^{30}$Si. The experimental data are taken from \cite{nucldata-30Si}$^{a}$ and \cite{Glatz80}$^{b}$
}
\begin{center}
\begin{tabular}{c|cc|cc}
\hline
& \multicolumn{2}{c}{exp.} &\multicolumn{2}{c}{cal.} \\
  &   	 $B(E2)$ &$M_p$ & 	$B(E2)$ &$M_p$ \\	
\hline
{Inband ($K=0^+_1$)}  	& & & & \\	
$	2^+_1	\rightarrow 0^+_1$	&	8.5$\pm$ 1.1$^{a}$	&	14	&	5.6	&	12.5	\\
$	4^+_1	\rightarrow 	2^+_1$	&	4.5 $^{+1.5}_{-1.0}$$^{b}$	&	14	 &	8.7	&	15.5	\\
\hline												
{Inband ($K=2^+_1$)}  	& & & & \\											
$	3^+_1	\rightarrow 	2^+_2$	&	7 $^{+16}_{-6}$$^{b}$	&	15	 &	8.5	&	15.4	\\
$	4^+_2	\rightarrow 	2^+_2$	&	4.1 $^{+3.7}_{-1.8}$$^{b}$	&	13	&	7.1	&	14	\\
$	4^+_2	\rightarrow 	3^+_1$	&		&		&		&	7.9	\\
$	5^+_1	\rightarrow 	3^+_1$	&	4.5 $^{+2.2}_{-1.4}$$^{b}$	&	15	&	7.6	&	14.5	\\
$	5^+_1	\rightarrow 	4^+2$	&	1.3 $^{+1.4}_{-0.7}$$^{b}$	&	8.1	&	4.5	&	11.1	\\
\hline												
{Inter-band}  			& & & & \\									
$	2^+_2	\rightarrow 	0^+_1$	&	1.7$\pm0.5$$^{a}$	&	5.9	&	1.6	&	6.6	\\
$	2^+_2	\rightarrow 	2^+_1$	&	8.9 $^{+11}_{-5}$$^{b}$	&	14	&	9.1	&	15.9	\\
$	3^+_1	\rightarrow 	2^+_1$	&	3.8 $^{+2.7}_{-1.3}$$^{b}$	&	11	&	2.8	&	8.8	\\
$	4^+_2	\rightarrow 	2^+_1$	&	7.8 $^{+4.1}_{-1.2}$$^{b}$	&	18	&	0.02	&	0.8	\\
$	4^+_1	\rightarrow 	3^+_1$	&		&		&	4.1	&	10.6	\\
\hline
& \multicolumn{2}{c}{exp.} &\multicolumn{2}{c}{cal.} \\
  &   	 $Q$ &$\mu$ & 		 $Q$ &$\mu$  \\	
  \hline
  $2^+_1$ &  $-5\pm 6$ & 0.76 & $-4.1$ & 1.3 \\
\hline
\end{tabular}
\end{center}
\end{table}

\begin{table}[ht]
\caption{\label{tab:mirror} The calculated and theoretical 
$B(E2)$ values and the proton transition matrix for the transitions 
$2^+_2\rightarrow 0^+_1$ and $2^+_1\rightarrow 0^+_1$ in $^{30}$Si and
$^{30}$S. The experimental values of the $M_n/M_p$ ratio are 
evaluated based on the mirror analysis
in which  $M_n$ of $^{30}$Si is assumed to be equal to $M_p$ of $^{30}$S.
The experimental $B(E2)$ values are taken from Ref.~\cite{nucldata-30Si}
and $^c$those evaluated from the life times and branching ratios.}
\begin{center}
\begin{tabular}{c|ccc|ccc}
\hline
 & \multicolumn{3}{c}{exp.} &\multicolumn{3}{c}{cal.} \\
  & $B(E2)$ &$M_p$ &$Mn/Mp$ & 	$B(E2)$ &$M_p$ &$Mn/Mp$ \\	
\hline
$2^+_1\rightarrow 0^+_1$ & & & & & & \\
$^{30}$Si &  8.5$\pm$1.1 & 15.3 & 1.1 & 5.6 & 12.5 & 1.2\\
$^{30}$S &  11$^c$ & 18 &  & 7.6 & 14.5 &  \\
\hline
$2^+_2\rightarrow 0^+_1$ & & & & & & \\
$^{30}$Si &  1.7$\pm$0.5 & 6.9 & 0.5 & 1.6 & 6.6 & 0.1 \\
$^{30}$S &  0.4$^c$        & 3   &     & 0.02 & 0.6 & \\ 
\hline
\end{tabular}
\end{center}
\end{table}

\subsection{$^{26}$Mg and $^{26}$Si}

In this subsection, we discuss the proton dominance in the $2^+_2\rightarrow 0^+_1$ of $^{26}$Si
in relation to the $\gamma$ deformation of the proton structure, 
in a similar way to the mirror analysis of $^{30}$Si and $^{30}$S.
Before discussing the $M_n/M_p$ ratio of $^{26}$Si based on the mirror analysis, 
we first investigate the structure of the ground and excited states of the mirror nucleus $^{26}$Mg
for which the existing data is richer than for $^{26}$Si.

\begin{figure}[th]
\epsfxsize=0.45\textwidth
\centerline{\epsffile{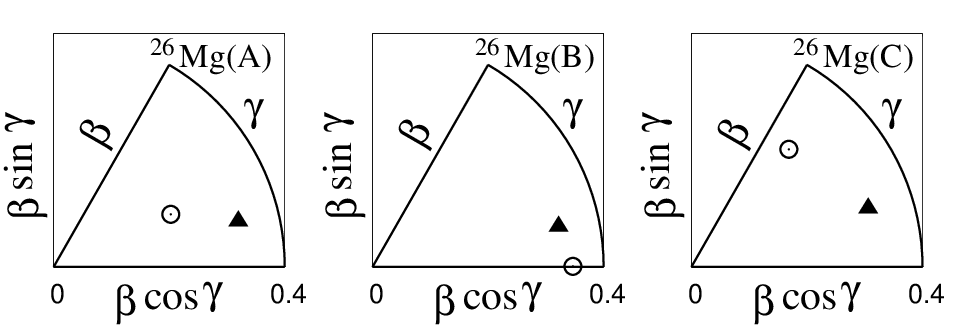}}
\caption{Deformation parameters $\beta$ and $\gamma$ of the intrinsic wave functions 
$^{26}$Mg(A), $^{26}$Mg(B), and $^{26}$Mg(C). The filled triangles indicate $\beta_p$ and $\gamma_p$ for the proton part
and the open circles are $\beta_n$ and $\gamma_n$ for the neutron part}
\label{fig:defo-mg26}
\end{figure}

We apply the AMD method to $^{26}$Mg. 
After energy variation for parity projected AMD wave functions, 
two local minimum solutions (A) and (B) are obtained in $^{26}$Mg, 
similarly to the case of $^{30}$Si.
The state (A) shows the triaxiality shape
$(\beta_p,\gamma_p)=(0.33, 0.08\pi)$ and
$(\beta_n,\gamma_n)=(0.22, 0.13\pi)$ with a smaller neutron deformation,
while the state (B) has a prolate deformation 
$(\beta_p,\gamma_p)=(0.33, 0.07\pi)$ and $(\beta_n,\gamma_n)=(0.35, 0.00\pi)$
with a larger neutron deformation  as seen in
Fig.~\ref{fig:defo-mg26}.  
The energy of the intrinsic state (A) degenerates with the state (B) within 0.1 MeV before the
angular momentum projection.
After the angular-momentum projection, 
the $0^+$ energy projected from the state (A)  
is 1.1 MeV higher than that obtained from the state (B).
In spite of the slightly higher energy of the state (A) than that of the state (B), we tentatively assign  
the states projected from (A) 
to the experimental $K^\pi=0^+$ ground band and its side band $K^\pi=2^+_1$, 
because the observed level structure and 
$E2$ transition strengths for the $K^\pi=0^+_1$ and $K^\pi=2^+_1$ band members can be reproduced by 
the results calculated with the state (A). The state (B) is inconsistent with the experimental fact that 
the $K^\pi=2^+_1$ side band exists by the ground band, and therefore, 
it would correspond to an excited $K^\pi=0^+$ band though it is eventually the lowest in the present calculations.  
In fact, the state (A) can be the lowest if we tune the interaction parameters, for instance, with $m=0.62$, $b=h=0.125$, and $u_{I}=-u_{II}=3200$ MeV. 
Hereafter, we concentrate only on the $K^\pi=0^+$ band and its $K^\pi=2^+$  side band constructed from the state (A),
and omit the mixing effect of the state (B). 

\begin{figure}[th]
\epsfxsize=0.25\textwidth
\centerline{\epsffile{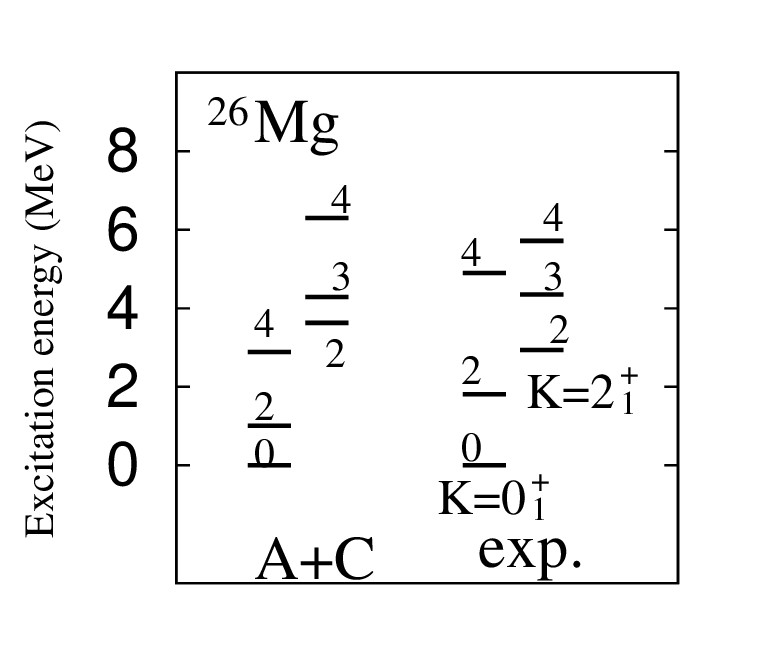}}
\caption{Energy levels of $^{26}$Mg. The calculated levels are obtained 
by the superposition of the states projected from the states (A) and (C).
The experimental levels are the positive parity bands
observed by the gamma-ray measurements in Ref.~\cite{Glatz86}.}
\label{fig:spe-mg26fig}
\end{figure}

To see the $\gamma$ softness of the $N=14$ neutron structure in $^{26}$Mg, we 
construct an intrinsic state (C) by the alternative energy variation that
we vary the centers of the single-particle Gaussian wave functions only for protons 
but freeze the neutron configuration  so that it has the same neutron structure as that
of the oblate ground state of $^{28}$Si. 
It means the energy variation with the constraint of 
the oblate neutron structure.
Thus obtained state (C) of $^{26}$Mg has an triaxial proton structure 
of $(\beta_p,\gamma_p)=(0.32, 0.10\pi)$
with the almost oblate neutron deformation $(\beta_n,\gamma_n)=(0.26, 0.28\pi)$.  
The state (C) is 2 MeV higher than the state (A) before and after the 
angular momentum. In Fig.~\ref{fig:spe-mg26fig}, we show the energy levels obtained by the superposition (A+C). 
The calculated energy levels are in reasonable agreements with the experimental data except for 
underestimation of the level spacing in the ground band. 
In case of $^{26}$Mg, the mixing of the state (C) gives minor contributions and 
the energy levels in the (A+C) calculations 
are qualitatively same as those obtained by the single intrinsic state (A)
without mixing of the state (C). It indicates that the $K^\pi=2^+_1$ band can be 
interpreted as the side band of the $K^\pi=0^+_1$ ground band constructed from the triaxial shape of 
the state (A).

Let us show the quadrupole transition properties calculated by the superposition (A+C)
and discuss the proton and neutron matrix amplitudes based on the mirror analysis. 
We first show the calculated values of the $B(E2)$ and $Q$ moments of $^{26}$Mg in Table \ref{tab:be2-mg26}. 
The present results are in reasonable agreement with the experimental data.
Next we discuss the $M_n/M_p$ ratio for the transitions $2^+_1\rightarrow 0^+_1$ and $2^+_2\rightarrow 0^+_1$
in $^{26}$Si. Here we assume the mirror symmetry that $M_n$ and $M_p$ of $^{26}$Si equal to $M_p$ and 
$M_n$ of $^{26}$Mg, respectively. The experimental values of $M_p$ are obtained from the $B(E2)$ values of 
$^{26}$Si and $^{26}$Mg. The experimental $M_n/M_p$ ratio for the ground-band transition 
$2^+_1\rightarrow 0^+_1$ is 0.94 which is close to one,  while that for the $2^+_2\rightarrow 0^+_1$ is 0.5.
The quenched $M_n/M_p$ for the transition $2^+_2\rightarrow 0^+_1$ indicates the proton dominance.
The present calculations reproduce this feature  of the proton dominance. 

Here we remind the reader that the $2^+_2$ state of  $^{26}$Mg is constructed 
from the triaxial neutron shape of the state (A). In other words, the $2^+_2$ state of 
the mirror nucleus $^{26}$Si is given by the rotation of the triaxial proton shape. 
Consequencely, the proton contribution is dominant 
in the excitation from the $0^+_1$ state to the $2^+_2$ state resulting in 
the small value of $M_n/M_p$ for the transition $2^+_2\rightarrow 0^+_1$ in $^{26}$Si.

\begin{table}[ht]
\caption{\label{tab:be2-mg26}
$E2$ transition strengths for the in-band and inter-band transitions
in $^{26}$Mg. The experimental data are taken from \cite{nucldata-26Mg}$^{a}$ and \cite{Glatz86}$^{b}$
}
\begin{center}
\begin{tabular}{c|cc|cc}
\hline
& \multicolumn{2}{c}{exp.} &\multicolumn{2}{c}{cal.} \\
 &   	 $B(E2)$ &$M_p$ & 	$B(E2)$ &$M_p$ \\	
\hline
{Inband ($K=0^+_1$)}  	& & & & \\												
$	2^+_1	\rightarrow	0^+_1	$ &	13.4$\pm$0.4$^{a}$	&	17.5 	&	12.9 	&	17.2 	\\
$	4^+_2	\rightarrow	2^+_1	$ &	14$\pm$3$^{a}$	&	24 	&	16.1 	&	25.7 	\\
\hline
{Inband ($K=2^+_1$)}  	& & & & \\														
$	3^+_2	\rightarrow		2^+_2	$ &	9.2 $^{+7.9}_{-4.5}$ $^{b}$	&	17 	&	22.5 	&	26.8 	\\
$	4^+_4	\rightarrow	3^+_2	$ &	5.2 $^{+6.1}_{-2.1}$ $^{b}$	&	15 	&	9.5 	&	19.8 	\\
\hline
{Inter-band}  			& & & & \\														
$	2^+_2	\rightarrow	0^+_1	$ &	0.35$\pm$0.07	$^{a}$	&	2.8 	&	0.1 	&	1.5 	\\
\hline
& \multicolumn{2}{c}{exp.} &\multicolumn{2}{c}{cal.} \\
  &   	 $Q$ &$\mu$ & 		 $Q$ &$\mu$  \\	
  \hline
  $2^+_1$ &  $-13.5\pm 2$ & 0.884 & $-14.2$ & 0.92 \\
  \hline
\end{tabular}
\end{center}
\end{table}

\begin{table}[ht]
\caption{\label{tab:mirror-26Si}
The calculated and theoretical 
$B(E2)$ values and the proton transition matrix for the transitions 
$2^+_2\rightarrow 0^+_1$ and $2^+_1\rightarrow 0^+_1$ in $^{26}$Si and
$^{26}$Mg. The experimental values of the $M_n/M_p$ ratio are 
evaluated based on the mirror analysis
in which  $M_n$ of $^{26}$Si is assumed to be equal to $M_p$ of $^{26}$Mg.
The experimental $B(E2)$ values are taken from Ref.~\cite{nucldata-26Mg}}
\begin{center}
\begin{tabular}{c|ccc|ccc}
\hline
 & \multicolumn{3}{c}{exp.} &\multicolumn{3}{c}{cal.} \\
 & $B(E2)$ &$M_p$ &$Mn/Mp$ & 	$B(E2)$ &$M_p$ &$Mn/Mp$ \\	
\hline
$2^+_1\rightarrow 0^+_1$ & & & & & & \\
$^{26}$Si &  15.4$\pm$1.5 & 18.8 & 0.94 & 6.4 & 12.1 & 1.4 \\
$^{26}$Mg &  13.4$\pm$0.4  & 17.5 &  & 12.9 & 17.2 &  \\
\hline
$2^+_2\rightarrow 0^+_1$ & & & & & & \\
$^{26}$Si &  1.6$\pm$0.5 & 6.0 & 0.5 & 1.4 & 5.6 & 0.26 \\
$^{26}$Mg &  0.35$\pm$0.07 & 2.8 &     & 0.09 & 1.5 &    \\ 
\hline
\end{tabular}
\end{center}
\end{table}

\begin{figure}[th]
\epsfxsize=0.4\textwidth
\centerline{\epsffile{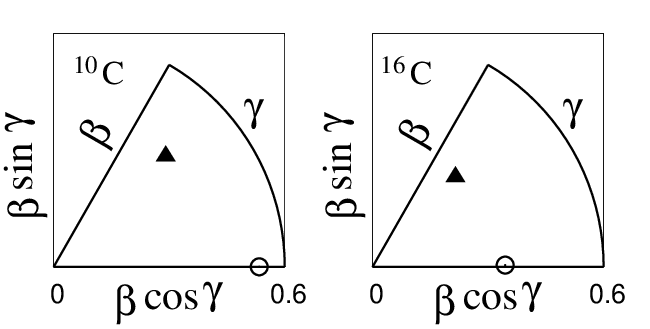}}
\caption{Deformation parameters for the intrinsic wave functions of $^{10}$C and $^{16}$C. 
The filled triangles indicate $\beta_p$ and $\gamma_p$ for the proton part
and the open circles are $\beta_n$ and $\gamma_n$ for the neutron part}
\label{fig:defo-c10-c16}
\end{figure}

\begin{figure}[th]
\epsfxsize=0.45\textwidth
\centerline{\epsffile{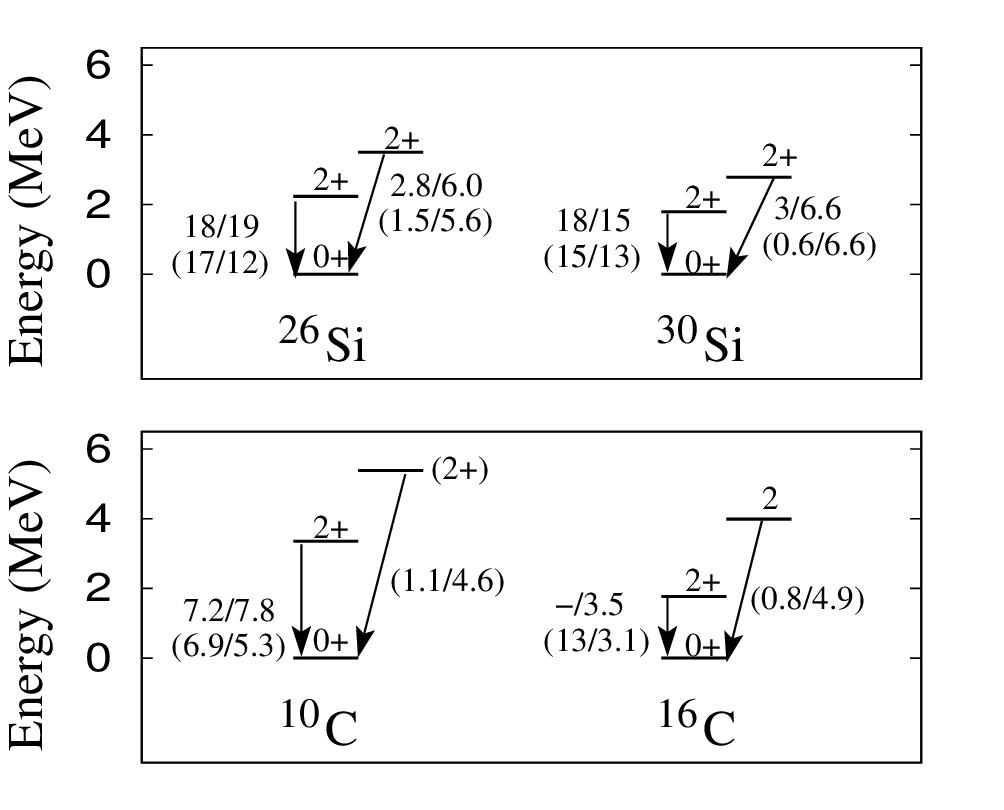}}
\caption{The experimental energy levels of the $0^+_1$, $2^+_1$, and $2^+_2$ states 
and $M_n/M_p$ ratios for the $2^+_1\rightarrow 0^+_1$ and $2^+_2\rightarrow 0^+_1$
of $^{26}$Si, $^{30}$Si, $^{10}$C and $^{16}$C\cite{nucldata-30Si,nucldata-26Mg,nucldata-10C,Ong:2007jb}. 
The experimental values of the neutron matrix amplitude($M_n$) are deduced from the corresponding 
$B(E2)$ values of the mirror nucleus.
The values in the parentheses are the present calculations.}
\label{fig:spe022}
\end{figure}
  
\section{$M_n/M_p$ ratios in $p$-shell nuclei} \label{sec:results2}
In the previous section, we discuss the 
the proton dominance in the transition $2^+_2\rightarrow 0^+_1$ 
between the side band and the ground band of 
Si isotopes with $N=Z\pm 2$. The proton dominance can be described by the 
triaxial shape of the proton structure. This feature
originates in the oblate trend of the $Z=14$ proton configuration.

Let us consider the proton dominance in $p$-shell nuclei. 
In C isotopes, $Z=6$ proton configuration has the oblate trend 
as is known in $^{12}$C. In fact, the oblate proton and prolate neutron shapes are
theoretically suggested in $^{10}$C and $^{16}$C 
in the AMD calculations\cite{KanadaEn'yo:2004bi}.
In these nuclei, the proton dominance
in the transition $2^+_2\rightarrow 0^+_1$ is expected 
as well as $^{26}$Si and $^{30}$Si.
In this section, we refer to the intrinsic structures of $^{10}$C and $^{16}$C
investigated in Ref.~\cite{KanadaEn'yo:2004bi}, 
and discuss the proton dominance in C isotopes in comparison with
Si isotopes. We use the AMD wave functions calculated in the previous work\cite{KanadaEn'yo:2004bi}.

We show the deformation parameters $\beta$ and $\gamma$ for proton and neutron density 
in the intrinsic states of $^{10}$C and $^{16}$C in Fig.~\ref{fig:defo-c10-c16}. The experimental and theoretical
values of the $M_n/M_p$ ratios are written in Fig.~\ref{fig:spe022}, in which the experimental 
energy levels are drawn.
In both nuclei, $^{10}$C and $^{16}$C, the proton structure shows the oblate deformation in spite of 
the prolate neutron structure. 
It means that the $Z=6$ proton structure is not so much affected by the neutron structure 
but it keeps the oblate tendency.
As a result, the decoupling between proton and neutron shapes 
is more remarkable in C isotopes than in Si isotopes. This decoupling is one of the reasons for
the enhanced $M_n/M_p$ ratio of the ground-band transition $2^+_1\rightarrow 0^+_1$
in $^{16}$C as discussed in Ref.~\cite{KanadaEn'yo:2004bi}.
For the transition $2^+_2\rightarrow 0^+_1$ in $^{10}$C and $^{16}$C, the calculated $M_n/M_p$ ratio is quenched 
and it indicates the proton dominance.
The suggested proton dominance originates in the oblate proton structure in the $Z=6$ systems. 
This is a good analogy to the proton dominance in $^{26}$Si and $^{30}$Si, which 
arises from the oblate trend of the $Z=14$ proton structure.

Experimentally, the $M_n/M_p$ ratio for $2^+_2\rightarrow 0^+_1$ is unknown. 
Inelastic scatterings of $^{10}$C and $^{16}$C might be a good probe to evaluate the 
$M_n/M_p$ ratio.

Note that the shape has large quantum fluctuation in such light systems and it is not observable.
Nevertheless, it is helpful to interpret the angular-momentum projected states 
in terms of rotation of the intrinsic deformation to get
semi-classical picture of transition properties.
Our argument is that the $M_n/M_p$ ratios in C isotopes can be qualitatively understood by the 
oblate proton and prolate neutron shapes.

\section{Summary}\label{sec:summary}
$E2$ transitions in  $^{30}$Si were investigated in relation with the  intrinsic deformation. 
In the calculation of the AMD method, 
the experimental $B(E2)$ values in the $K^\pi=0^+_1$ and $K^\pi=2^+_1$ bands 
are reproduced by the calculation. 
Based on mirror analysis, the transition matrix amplitudes $M_p$ and $M_n$ 
were discussed. 
Particular attention is paid to the 
$M_n/M_p$ ratio in the transition from the $2^+_2$ state to the $0^+_1$ state, 
whose quenching is experimentally known.
The $M_n/M_p$ ratio in $^{26}$Mg and $^{26}$Si was also investigated.  
We have shown that the $K^\pi=2^+_1$ band can be interpreted as the triaxial side band, 
and the proton contribution is dominant in the transition $2^+_2\rightarrow 0^+_1$.
The quenched $M_n/M_p$, i.e., the proton dominance in $2^+_2\rightarrow 0^+_1$
in $^{30}$Si and $^{26}$Si originates in the oblate trend of the $Z=14$ proton structure.
The proton dominance in $2^+_2\rightarrow 0^+_1$ is suggested also in $^{10}$C and $^{16}$C, 
where the oblate proton structure is favored. 

We should comment that the shape has large quantum fluctuation 
in light-mass systems and it is not observable. 
Strictly speaking, the macroscopic picture may be too simple for such systems, and therefore, 
it is not easy to clearly distinguish between two collective pictures, 
the rotational mode of a static triaxial state and the $\gamma$-vibrational mode. 
Nevertheless, it is helpful to interpret the angular-momentum projected states 
in terms of rotation of the intrinsic deformation to get
semi-classical picture of transition properties.
Our argument is that the quesched $M_n/M_p$ ratios in $2^+_2\rightarrow 0^+_1$ of these nuclei 
can be qualitatively understood by the oblate trend of proton shapes in the prolate neutron structures.

\section*{Acknowledgments}
The computational calculations of this work were performed by using the
supercomputers at YITP and done in Supercomputer Projects 
of High Energy Accelerator Research Organization (KEK).
This work was supported by Grant-in-Aid for Scientific Research from Japan Society for the Promotion of Science (JSPS).
It was also supported by 
the Grant-in-Aid for the Global COE Program "The Next Generation of Physics, 
Spun from Universality and Emergence" from the Ministry of Education, Culture, Sports, Science and Technology (MEXT) of Japan.

\end{document}